\documentclass[aps,prb,reprint,showpacs]{revtex4-1}
\usepackage{graphicx}

\begin{document}

\title{Properties of BaTiO$_3$/BaZrO$_3$ ferroelectric superlattices with
competing instabilities}

\author{Alexander I. Lebedev}
\email[]{swan@scon155.phys.msu.ru}
\affiliation{Physics Department, Moscow State University, \\
Leninskie gory, 119991 Moscow, Russia}

\date{\today}

\begin{abstract}
Properties of (BaTiO$_3$)$_1$/(BaZrO$_3$)$_n$ ferroelectric superlattices (SLs)
with $n = {}$1--7 grown in the [001] direction are calculated from first
principles within the density functional theory. It is revealed that the
quasi-two-dimensional ferroelectricity occurs in these SLs in the barium
titanate layers with a thickness of one unit cell; the polarization is oriented
in the layer plane and weakly interacts with the polarization in neighboring
layers. The ferroelectric ordering energy and the height of the barrier
separating different orientational states of polarization in these SLs are
sufficiently large to provide the formation of an array of independent
polarized planes at 300~K. The effect of the structural instability on the
properties of SLs is considered. It is shown that the ground state is a result
of simultaneous condensation of the $\Gamma_{15}$ polar phonon and phonons
at the $M$ point (for SLs with even period) or at the $A$ point (for SLs with
odd period); it is a polar structure with out-of-phase rotations of the
octahedra in neighboring layers, in which highly polarized layers are
spatially separated from the layers with strong rotations. The competition
between the ferroelectric and structural instabilities in biaxially compressed
SLs manifests itself in that the switching on of the octahedra rotations leads
to an abrupt change of the polarization direction and can cause an improper
ferroelectric phase transition to occur. It was shown that the experimentally
observed $z$-component of polarization in the SLs can appear only as a result
of the mechanical stress relaxation.
\end{abstract}


\maketitle

\section{Introduction}

Ferroelectric superlattices (SLs) represent a new class of artificial materials
whose properties, such as the Curie temperature and spontaneous polarization,
often exceed those of bulk ferroelectrics.~\cite{CurrOpinSolidStateMaterSci.9.122,
Ghosez-Junquera}  An analysis of properties of ten ferroelectric SLs with the
perovskite structure grown in the [001] direction have shown that they are
ferroelectrics at low temperatures, no matter whether they are composed of two
ferroelectrics, a ferroelectric and a paraelectric, or even two
paraelectrics.~\cite{PhysSolidState.52.1448}  One of the possible applications
of SLs can be non-volatile ferroelectric random access memory (FRAM) devices
in electronics.

The recent studies of the ferroelectric instability in free-standing
(KNbO$_3$)$_1$(KTaO$_3$)$_n$ superlattices grown in the [001] direction showed
that, with increasing $n$, the tendency to the ferroelectric ordering is
retained in KNbO$_3$ layers with a thickness of one unit cell while the
interaction energy between neighboring KNbO$_3$ layers exponentially
decreases.~\cite{PhysSolidState.53.2463,PhysStatusSolidiB.249.789}  It was shown
that at $n \ge 2$ the ground state of the SLs is an array of ferroelectrically
polarized planes whose polarization is concentrated in the KNbO$_3$ layer, is
parallel to the [110] direction, and weakly interacts with polarization in
neighboring layers. The use of such arrays of quasi-two-dimensional ferroelectric
planes as a medium for the three-dimensional data recording can make it possible
to achieve a volume recording density of $\sim$$10^{18}$\,bit/cm$^3$.

Unfortunately, the implementation of the above-described ground state in the
(KNbO$_3$)$_1$(KTaO$_3$)$_n$ SLs requires very low temperatures ($\sim$4~K)
because of the low ferroelectric ordering energy and low height of the potential
barrier separating different orientational states of polarization. That is why
one of the aims of this work was to study the feasibility of similar
quasi-two-dimensional structures and the ways to improve their characteristics
in BaTiO$_3$/BaZrO$_3$ ferroelectric superlattices with an active layer of
barium titanate.

In BaTiO$_3$/BaZrO$_3$ superlattices, the tension of the BaTiO$_3$ layers as
a result of the epitaxial matching with the BaZrO$_3$ layers having a larger
lattice parameter should stabilize the $Amm2$ ground state necessary to
implement the quasi-two-dimensional structure~\cite{PhysSolidState.53.2463}
in the BaTiO$_3$ layers. However, since the simultaneous compression of the
BaZrO$_3$ layers can cause the polarization normal to the interface to occur
in them,~\cite{PhysRevB.72.144101} the search for the ground-state structure
in these SLs requires a detailed analysis. Furthermore, the instability of
the BaZrO$_3$ structure with respect to the octahedra
rotation~\cite{PhysRevB.72.205104,PhysRevB.73.180102,PhysRevB.79.174107}
can result in the competition of the ferroelectric and structural instabilities
in the SLs, and this problem also calls for a detailed study.

The properties of BaTiO$_3$/BaZrO$_3$ SLs have been studied previously both
experimentally~\cite{JApplPhys.91.2284,ThinSolidFilms.509.13,
Ferroelectrics.346.56, ApplPhysLett.92.102903, JApplPhys.104.114105,
JApplPhys.108.084104} and theoretically.~\cite{PhysSolidState.52.1448}  The
initial interest to these SLs was caused by giant dielectric constants observed
in them.~\cite{JApplPhys.91.2284}  Room-temperature observations of
ferroelectric hysteresis loops in the SLs~\cite{ThinSolidFilms.509.13,
ApplPhysLett.92.102903,JApplPhys.108.084104}  suggested that the ferroelectric
transition temperature is them exceeds 300~K. The spontaneous polarization
nonmonotonically varied with the period of SLs and depended on the electrode
configuration (in short-period SLs with electrodes on both sides of the film,
hysteresis loops were absent~\cite{ApplPhysLett.92.102903}). These features
have already been discussed in Ref.~\onlinecite{PhysSolidState.52.1448} and
were explained by the rotation of the polarization to the layer plane. The
recent study of Raman spectra and dielectric properties of BaTiO$_3$/BaZrO$_3$
SLs revealed the appearance of the $z$-component of polarization in barium
titanate layers in the SLs grown on MgO substrates.~\cite{JApplPhys.108.084104}
Interpretation of these results requires the study of the influence of
substrate-induced strain on the ground-state structure of the SLs.

\section{Calculation technique}

The (BaTiO$_3$)$_1$(BaZrO$_3$)$_n$ superlattices considered in this work are
periodic structures grown in the [001] direction and consisting of the BaTiO$_3$
layer with a thickness of one unit cell and the BaZrO$_3$ layer with a thickness
of $n$ unit cells ($1 \le n \le 7$). These SLs were modeled on supercells of
1$\times$1$\times$L unit cells, where $L = n + 1$ is the SL period; in modeling
the structures generated by instabilities at the $M$ and $A$ points at the
boundary of the Brillouin zone, the primitive cell volume was doubled. In this
work, particular attention was paid to free-standing SLs. In addition, a number
of calculations were performed for biaxially compressed (BaTiO$_3$)$_1$(BaZrO$_3$)$_1$
and (BaTiO$_3$)$_2$(BaZrO$_3$)$_2$ SLs and for free-standing
(BaTiO$_3$)$_n$(BaZrO$_3$)$_n$ SLs with $n = 2$, 3, and 4.

The calculations were performed using the density functional theory with
pseudopotentials and wave function expansion in plane waves as implemented in
the ABINIT code.~\cite{abinit3}  As in the earlier study of these
SLs,~\cite{PhysSolidState.52.1448}  the exchange--correlation interaction was
described in the local density approximation (LDA). As pseudopotentials, the
optimized separable nonlocal pseudopotentials constructed using the OPIUM
program~\cite{opium} were used; the local potential correction was added to
them to improve the transferability. The parameters used for constructing
the pseudopotentials and other calculation details are given in
Refs.~\onlinecite{PhysSolidState.51.362,PhysSolidState.52.1448}. The cut-off
energy for the plane waves was 30~Ha (816~eV); the integration over the
Brillouin zone was performed using the 8$\times$8$\times$4 Monkhorst--Pack
mesh for SLs with $n = 1$ and 2 and the 8$\times$8$\times$2 one for SLs with
$n = {}$3--7. The relaxation of the lattice parameters and atomic positions
was performed until the Hellmann--Feynman forces became less than
$5 \cdot 10^{-6}$~Ha/Bohr (0.25\,meV/{\AA}). In the case of biaxially compressed
SLs, the in-plane lattice parameter was fixed equal to 0.97$a_0$, 0.98$a_0$,
and 0.99$a_0$, where $a_0 = 7.7182$~Bohr (4.0843~{\AA}) is the in-plane lattice
parameter for the $P4/mmm$ phase of the free-standing
(BaTiO$_3$)$_1$(BaZrO$_3$)$_1$ SL. The phonon spectra were calculated using
the equations obtained from the density functional perturbation theory. The
total spontaneous polarization in the SLs was calculated by the Berry phase
method; its distribution over layers was calculated using the formula
$P_{\alpha} = \sum_i w_i Z^*_{i, \alpha \beta} u_{i, \beta}$ from
displacements $u_i$ of atoms obtained in the polar phase and tensors of their
effective Born charges $Z^*_{i, \alpha \beta}$ in the paraelectric phase;
here $w_i$ are the weight factors equal to unity for atoms in the Ti--O or
Zr--O layer under consideration, 1/2 for atoms in neighboring Ba--O layers,
and zero for other atoms.

The properties of BaTiO$_3$ obtained using the described approach were
published previously~\cite{PhysSolidState.51.362}  and are in good agreement
with experiment.

\section{Results}

\subsection{Structure of BaZrO$_3$}
\label{sec31}

\begin{figure}
\centering
\includegraphics{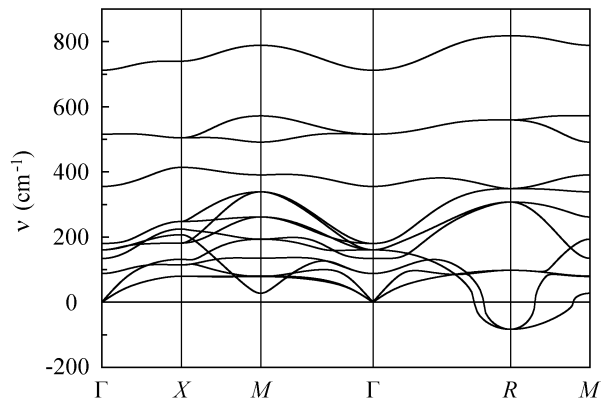}
\caption{\label{fig1}The phonon spectrum of BaZrO$_3$ in the cubic phase.}
\end{figure}

The structure of BaZrO$_3$ calculated using the described approach is in good
agreement with experiment and results of earlier
calculations.~\cite{PhysRevB.72.205104,PhysRevB.73.180102,PhysRevB.79.174107}
For example, the lattice parameter in cubic BaZrO$_3$ is 7.8724~Bohr
(4.1659~{\AA}) and differs from the experimental value (4.191~{\AA} at 2~K,
Ref.~\onlinecite{PhysRevB.72.205104}) by 0.60\% (the small underestimate of the
lattice parameters is characteristic of the used LDA approximation). As in earlier
studies,~\cite{PhysRevB.72.205104,PhysRevB.73.180102,PhysRevB.79.174107}  the
calculations of the phonon spectrum of cubic BaZrO$_3$ reveal an unstable phonon
mode at the $R$ point at the boundary of the Brillouin zone with a frequency of
81$i$~cm$^{-1}$ (Fig.~\ref{fig1}). The unstable phonon at the $M$ point observed
in Ref.~\onlinecite{PhysRevB.79.174107} is absent in our calculations.

However, the instability of the phonon spectrum of BaZrO$_3$ at the $R$ point
has not yet been confirmed experimentally. One of the possible causes of
disagreement between experiment and theory can be quantum
fluctuations~\cite{PhysRevB.72.205104} which can destroy the long-range order
in rotations of the oxygen octahedra.%
    \footnote{EXAFS measurements of the Debye--Waller factor for Ba--O bonds
    in BaZrO$_3$ revealed its anomalously high values at 300~K, corresponding
    to the amplitude of the local rotations of $\sim$4~degrees. In our opinion,
    this gives an evidence of the instability under consideration. The results
    of these studies will be published elsewhere.}
According to our calculations, the ground-state structure of BaZrO$_3$ is
$I4/mcm$. This result differs from predictions of Ref.~\onlinecite{PhysRevB.73.180102},
in which the $P{\bar 1}$ structure was considered to be equilibrium. In the
present study, it was shown that the $P{\bar 1}$ structure is nonequilibrium
and slowly relaxes to the $I4/mcm$ structure. The calculated static dielectric
constant of barium zirconate in the ground state is
$\varepsilon_{xx} = \varepsilon_{yy} = 58.8$, $\varepsilon_{zz} = 53.3$ (the
experimental value is 47, Ref.~\onlinecite{PhysRevB.72.205104}).

\subsection{The BaTiO$_3$/BaZrO$_3$ superlattices}

\begin{figure}
\centering
\includegraphics{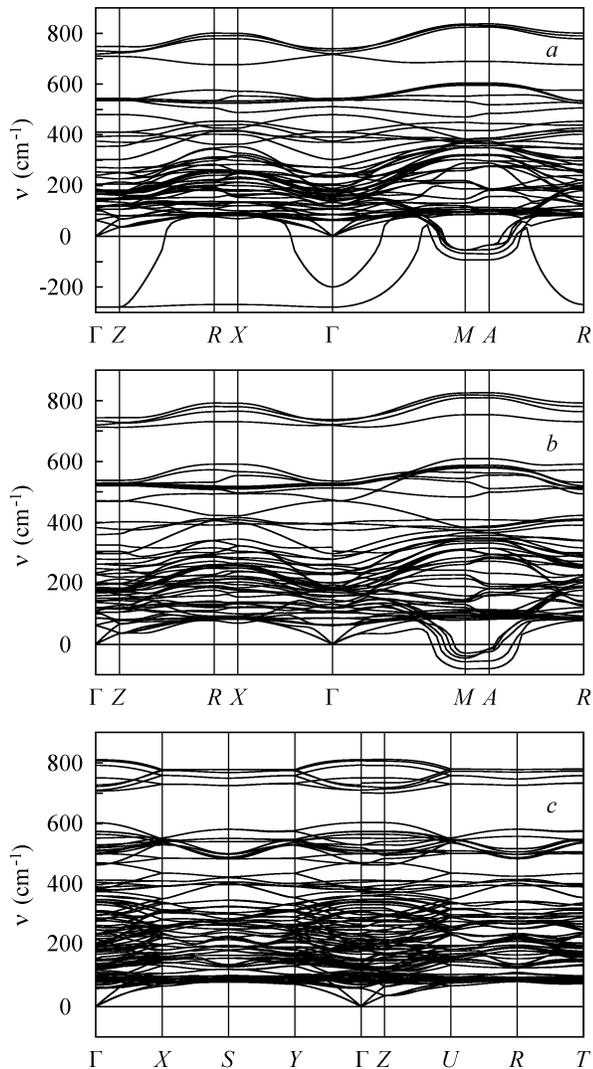}
\caption{\label{fig2}Phonon spectra of the free-standing
(BaTiO$_3$)$_1$(BaZrO$_3$)$_3$ superlattice in (\emph{a}) $P4/mmm$, (\emph{b})
$Amm2$, and (\emph{c}) $Pmc2_1$ phases.}
\end{figure}

The phonon spectra of different phases of the free-standing
(BaTiO$_3$)$_1$(BaZrO$_3$)$_3$ superlattice are shown in Fig.~\ref{fig2}. One
can see that two types of instability are observed in the phonon spectrum of
the paraelectric $P4/mmm$ phase (Fig.~\ref{fig2},\emph{a}): the ferroelectric
instability and the structural instability associated with the oxygen octahedra
rotation. The phonon frequencies at high-symmetry points of the Brillouin zone
in the paraelectric phase of superlattices with $n = 1$, 2, and 3 are given
in Table~\ref{table1}.

\begin{table*}
\caption{\label{table1}The lowest frequencies of optical phonons at
high-symmetry points of the Brillouin zone in $P4/mmm$, $Amm2$ phases,
and in the ground state ($Pmc2_1$) of free-standing
(BaTiO$_3$)$_1$(BaZrO$_3$)$_n$ superlattices with $n = 1$, 2, and 3.}
\begin{ruledtabular}
\begin{tabular}{ccccccc}
$n$ & \multicolumn{6}{c}{Phonon frequencies (cm$^{-1})$} \\
\cline{2-7}
    & $\Gamma$~(0,0,0) & $Z$~(0,0,$\frac{1}{2}$) & $X$~($\frac{1}{2}$,0,0) & $R$~($\frac{1}{2}$,0,$\frac{1}{2}$) & $M$~($\frac{1}{2}$,$\frac{1}{2}$,0) & $A$~($\frac{1}{2}$,$\frac{1}{2}$,$\frac{1}{2}$) \\
\hline
    & \multicolumn{6}{c}{The $P4/mmm$ phase} \\
1   & 248$i$ & 243$i$ & 233$i$ & 228$i$ & 90$i$                      & 86$i$ \\
2   & 269$i$ & 268$i$ & 257$i$ & 257$i$ & 88$i$, 65$i$               & 90$i$, 63$i$, 21$i$ \\
3   & 279$i$ & 279$i$ & 268$i$ & 268$i$ & 92$i$, 66$i$, 54$i$, 53$i$ & 91$i$, 68$i$, 51$i$, 34$i$ \\
\hline
    & \multicolumn{6}{c}{The $Amm2$ phase} \\
1   &  91    &  66    &  87    &  89    & 60$i$                             & 54$i$ \\
2   &  76    &  46    &  84    &  79    & 69$i$, 39$i$                      & 72$i$, 41$i$ \\
3   &  61    &  36    &  68    &  77    & 79$i$, 58$i$, 46$i$, 38$i$, 25$i$ & 79$i$, 53$i$, 20$i$, 16$i$ \\
\hline
 & $\Gamma$~(0,0,0) & $Z$~(0,0,$\frac{1}{2}$) & $X$/$Y$ & $U$/$T$ & $S$ & $R$ \\
\hline
    & \multicolumn{6}{c}{The $Pmc2_1$ phase} \\
1   &  84    &  66    & 70/76  & 78/82  & 86                                & 89 \\
3   &  60    &  35    & 67/73  & 68/74  & 80                                & 81 \\
\end{tabular}
\end{ruledtabular}
\end{table*}

The instability region appearing as a band of imaginary phonon frequencies
along the path $\Gamma$--$Z$--$R$--$X$--$\Gamma$ (imaginary frequencies are
presented as negative numbers in the figure) is caused by the ferroelectric
instability of ...--Ti--O--... chains, which was established for the first time
Ref.~\onlinecite{PhysRevLett.74.4067}.  An analysis of the eigenvectors of
phonons related to this instability region shows that out-of-phase transverse
Ti and O atomic displacements in the $xy$ plane in chains propagating in the
[100] and [010] directions dominate in these eigenvectors at all points of
the Brillouin zone; at the Brillouin zone center, this displacement pattern
corresponds to the doubly degenerate ferroelectric $E_u$ mode. For out-of-phase
atomic displacements in the chains propagating in the [001] direction and
consisting of alternating titanium and zirconium atoms (...--Ti--O--Zr--O--...),
the lowest-energy ferroelectric $A_{2u}$ mode at the $\Gamma$ point is always
stable; its frequency increases from 37~cm$^{-1}$ for $n = 1$ to 85~cm$^{-1}$
for $n = 3$. A similar instability region along the path $\Gamma$--$Z$--$R$--$X$--$\Gamma$
was observed in KNbO$_3$/KTaO$_3$ superlattices.~\cite{PhysStatusSolidiB.249.789}

Among the possible polar phases resulting from the condensation of the unstable
$E_u$ mode at the $\Gamma$ point, the $Amm2$ phase with polarization along
the [110] direction has the lowest energy in (BaTiO$_3$)$_1$(BaZrO$_3$)$_n$
SLs with $n = {}$1--7. However, the phonon spectra of this phase
(Fig.~\ref{fig2},\emph{b}) show that phonons at the $M$ and $A$ points at the
boundary of the Brillouin zone remain unstable, which indicates the
instability of the $Amm2$ structure with respect to the oxygen octahedra
rotation.%
    \footnote{Hereafter, we shall use the notation of the points in the
    Brillouin zone of the tetragonal paraelectric structure for their notation
    in the low-symmetry structures.}
This instability of the polar $Amm2$ phases is characteristic of SLs with all
periods (see Table~\ref{table1}). Thus, the $Amm2$ phase is not a true
ground-state structure for the SLs under consideration.

As follows from Table~\ref{table1}, the frequencies of two unstable phonons
at the $M$ and $A$ points for the $Amm2$ phase of the superlattices with different
periods are very close. Therefore, to search for the true ground state, it is
necessary to compare the energies of all structures resulting from the condensation
of these two phonons.

\begin{figure}
\centering
\includegraphics{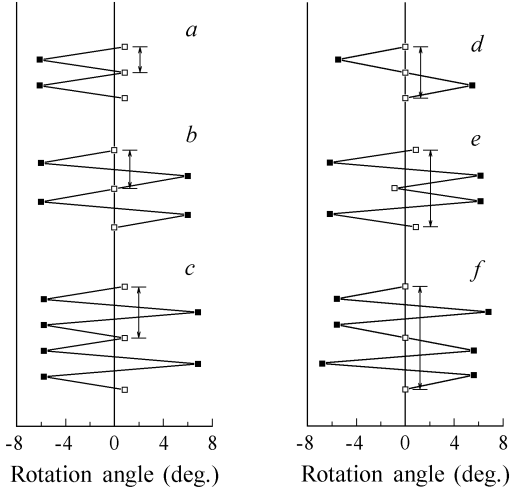}
\caption{\label{fig3}The octahedra rotation angles in the ground state
of (BaTiO$_3$)$_1$(BaZrO$_3$)$_n$ superlattices corresponding to the
condensation of phonons at the (\emph{a--c}) $M$ and (\emph{d--f}) $A$ points.
(\emph{a,d}) $n = 1$ ($L = 2$), (\emph{b,e}) $n = 2$ ($L = 3$),
(\emph{c,f}) $n = 3$ ($L = 4$). The BaTiO$_3$ and BaZrO$_3$ layers are
denoted by open and filled squares, respectively. Vertical lines with arrows
indicate the physical periods of the resulting structures.}
\end{figure}

An analysis of the eigenvectors of unstable phonons at the $M$ point shows
that the most unstable phonons are characterized by out-of-phase rotations
of octahedra in neighboring layers. In SLs with even period ($L = 2$, 4),
the rotation angle is small in the BaTiO$_3$ layers and large in the
BaZrO$_3$ layers (Figs.~\ref{fig3},\emph{a} and \ref{fig3},\emph{c}); in SLs
with odd period ($L = 3$), the rotation angle is zero in the BaTiO$_3$ layers
and is large in the BaZrO$_3$ layers (Fig.~\ref{fig3},\emph{b}). Modes with
lower instability (Table~\ref{table1}) include modes with other combinations
of rotations in layers (in particular, in-phase rotations in BaZrO$_3$ layers
and strong out-of-phase rotations in every second BaZrO$_3$ layer) and a
doubly degenerate mode with a complex combination of distortion and tilting
of the oxygen octahedra (for $L = 4$).

An analysis of the eigenvectors of unstable phonons at the $A$ point shows
that in SLs with even period ($L = 2$, 4), the most unstable phonons are
characterized by the out-of-phase rotations in neighboring BaZrO$_3$ layers
and the absence of rotations in the BaTiO$_3$ layers (Figs.~\ref{fig3},\emph{d}
and \ref{fig3},\emph{f}). In SLs with odd period ($L = 3$), out-of-phase
rotation angles are large in the BaZrO$_3$ layers and small in the BaTiO$_3$
layers (Fig.~\ref{fig3},\emph{e}). In both cases, the \emph{physical period}
of SLs is doubled in comparison with the period $L$ specified by the layer
sequence. Modes with lower instability (Table~\ref{table1}) are qualitatively
similar to the above-described modes at the $M$ point.

An abrupt change in the rotation angle in going from the BaTiO$_3$ layer to
the BaZrO$_3$ layer and its weak dependence on the period of SL correlate
with the small change of the unstable phonon frequency along the $M$--$A$ line,
which indicates a strong localization of these rotations in the layers.

\begin{table}
\caption{\label{table2}The lattice parameters in the paraelectric $P4/mmm$
phase of free-standing (BaTiO$_3$)$_1$(BaZrO$_3$)$_n$ superlattices with
$1 \le n \le 7$, the energy of the ferroelectric ground state $\Delta E_1$
(the $Amm2$ phase), the energy $\Delta E_2$ of the $I4/mcm$ or $P4/mbm$ phases
with pure rotations, the energy of the true ground state
$\Delta E_3$ ($Ima2$ or $Pmc2_1$ phase), and the polarization in the
ferroelectric ground ($P_{s1}$) and true ground ($P_{s3}$) states.}
\begin{ruledtabular}
\begin{tabular}{ccccccc}
$n$                  & 1       & 2        & 3        & 4        & 5        & 7 \\
\hline
$a_0$ (Bohr)         & 7.7182  & 7.7747   & 7.8009   & 7.8160   & 7.8258   & 7.8379 \\
$c_0$ (Bohr)         & 15.3298 & 23.1923  & 31.0604  & 38.9304  & 46.8013  & 62.5443 \\
$\Delta E_1$ (meV)   & $-$72.7 & $-$101.0 & $-$115.1 & $-$123.6 & $-$129.3 & $-$136.2 \\
$\Delta E_2$ (meV)   & $-$21.3 & $-$34.6  & $-$46.7  & $-$57.7  & $-$68.3  & $-$89.2 \\
$\Delta E_3$ (meV)   & $-$78.3 & $-$116.8 & $-$142.3 & $-$161.5 & $-$177.8 & $-$205.3 \\
$P_{s1}$ (C/m$^2$)   & 0.2755  & 0.2024   & 0.1568   & 0.1273   & 0.1070   & --- \\
$P_{s3}$ (C/m$^2$)   & 0.2591  & 0.1859   & 0.1445   & 0.1113   & 0.0984   & --- \\
\end{tabular}
\end{ruledtabular}
\end{table}

To search for the ground state of SLs, the oxygen octahedra rotations
corresponding to the least stable phonons at the $M$ and $A$ points of the
Brillouin zone were added to the polar $Amm2$ structure. A comparison of the
total energies of the structures resulting from simultaneous condensation of
the ``rotational'' and ferroelectric modes shows that the ground-state
structure depends on the superlattice period $L$: for odd $L$, it is described
by the space group $Ima2$ and results from the condensation of the unstable
``rotational'' mode at the $A$ point; for even $L$, it is described by the
space group $Pmc2_1$ and results from the condensation of the unstable
``rotational'' mode at the $M$ point. The energy difference of the structures
resulting from the condensation of phonons at the $M$ and $A$ points is rather
small (1.6--1.9~meV). The energies of the structures corresponding to the
ferroelectric ground state and to the true ground state, as well as to nonpolar
structures with pure octahedra rotations (space group $I4/mcm$ for odd $L$ and
$P4/mbm$ for even $L$) are given in Table~\ref{table2}.%
    \footnote{The ferroelectric ground state is a structure with a minimum
    energy obtained by taking into account only the ferroelectric unstable
    mode.}
An increase in all three energies with increasing $n$ is due to two effects:
(i) an increase in the BaZrO$_3$ volume fraction in the structure, which results
in an increase of the instability of SLs with respect to the octahedra rotation;
(ii) an increase in biaxial tension of the BaTiO$_3$ layers (as a consequence
of an increase in the in-plane lattice parameter), which enhances the
ferroelectric instability. The corresponding change in the frequencies of unstable
phonons at the $\Gamma$ point can be seen in Table~\ref{table1}.

\begin{figure}
\centering
\includegraphics{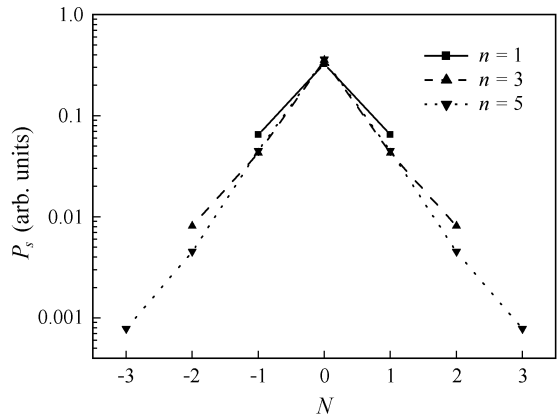}
\caption{\label{fig4}The polarization distribution between the layers in the
ground-state structure of (BaTiO$_3$)$_1$(BaZrO$_3$)$_n$ superlattices with
$n = 1$, 3, and 5. The layer number $N = 0$ corresponds to the BaTiO$_3$ layer.}
\end{figure}

The polarization profiles in the obtained ground states for SLs with $n = 1$,
3, and 5 are shown in Fig.~\ref{fig4}. One can see that the polarization is
concentrated in the barium titanate layer and almost exponentially decreases
with distance with a characteristic length scale of $\sim$2.0~{\AA}. It is
interesting that the spontaneous polarization in the true ground state is
lower than the polarization in the ferroelectric ground state by only 6--12\%
(Table~\ref{table2}). This indicates the weak effect of the octahedra rotations
on the spontaneous polarization in SLs.%
    \footnote{An additional cause of the decrease in the polarization in
    free-standing SLs when the octahedra rotations are switched on can be
    a systematic decrease (by 0.02--0.03~Bohr) in the in-plane lattice parameter.}
We explain this by the fact that regions with high polarization in the SLs under
consideration are spatially separated from regions with strong octahedra rotations,
and thus the competition between the structural and ferroelectric instabilities
becomes significantly decreased. This feature essentially distinguishes the
superlattices from such crystals as CaTiO$_3$ and PbTiO$_3$ with competing
instabilities in the cubic phase. In the former case, the structural distortions
completely suppress ferroelectricity; in the latter case, an opposite effect occurs.

To test the possibility of forming the arrays of quasi-two-dimensional polarized
planes in the ground state of (BaTiO$_3$)$_1$(BaZrO$_3$)$_n$ SLs, we estimated
the interlayer interaction energy. The energy 2$W_{\rm int}$ (the energy
corresponding to two domain walls) was calculated as the difference of the
total energies of ferroelectrically and antiferroelectrically ordered SLs
with a doubled period in a similar way as in
Ref.~\onlinecite{PhysSolidState.53.2463}; in this calculation, the octahedra
rotations were neglected. The value of 2$W_{\rm int}$ was 7.46~meV for
$n = 1$, 1.505~meV for $n = 2$, and 0.420~meV for $n = 3$, and decreased
almost exponentially with increasing $n$. For all $n$, this energy was much
lower than the energy gain resulting from the ferroelectric ordering
$\Delta E_1$ (Table~\ref{table2}), which ensures almost independent
polarizations in the quasi-two-dimensional layers and the formation of the
arrays of independent polarized planes similar to those observed in
Ref.~\onlinecite{PhysSolidState.53.2463}. The potential barrier height
$\Delta U$ for reorientation of the polarization, which occurs in the SLs
under consideration by rotating the polarization vector in the layer plane,
was 22.9~meV (per Ti atom) for $n = 1$, 33.2~meV for $n = 2$, and 38.3~meV
for $n = 3$. The obtained potential barrier heights and ferroelectric
ordering energies are much higher than those for the KNbO$_3$/KTaO$_3$ SLs
considered in Refs.~\onlinecite{PhysSolidState.53.2463,PhysStatusSolidiB.249.789}.
In our opinion, these values are sufficiently large to provide stable
polarization in the arrays of quasi-two-dimensional polarized planes at 300~K.

\section{Discussion}

\begin{figure}
\centering
\includegraphics{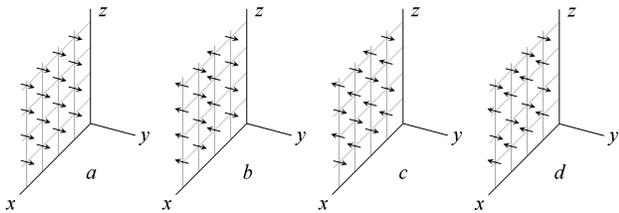}
\caption{\label{fig5}Patterns of the polarization ordering in the
...--Ti--O--... chains propagating along the $y$ axis which accompany the
condensation of unstable phonon at the (\emph{a}) $\Gamma$, (\emph{b}) $X$,
(\emph{c}) $R$, and (\emph{d}) $Z$ points.}
\end{figure}

We start the discussion with an analysis of possible polar structures in
(BaTiO$_3$)$_1$(BaZrO$_3$)$_n$ SLs. The ferroelectric instability in the
...--Ti--O--... chains can lead not only to the ferroelectric ordering, but
also to the formation of antiferroelectric structures resulting from the
condensation of unstable phonons at $Z$, $R$, and $X$ points at the boundary
of the Brillouin zone. Three types of the polarization ordering in the chains,
which result from the condensation of these unstable phonons in the paraelectric
$P4/mmm$ phase, are shown in Fig.~\ref{fig5}. It is seen that these structures
differ only in the relative orientation of polarization in neighboring chains.
For example, for the phonon condensation at the $X$ point, the antiferroelectric
$Pmma$ phase appears in the SLs; the modulation wave vector in this phase is
directed along the $x$ axis, whereas the out-of-phase displacements of Ti and
O atoms forming the chains are directed along the $y$ axis (Fig.~\ref{fig5},\emph{b}).
It is remarkable that this phase still exhibits the ferroelectric instability
with respect to the polar displacements along the $x$ axis and finally
transforms to the polar $Pm$ ($Pmc2_1$ for $n = 1$) phase being \emph{ferrielectric}.
Similarly, the antiferroelectric $Cmma$ phase with the modulation described by
the phonon at the $R$ point and with out-of-phase displacements along the $y$
axis (Fig.~\ref{fig5},\emph{c}) also is ferroelectrically unstable and finally
transforms to the \emph{ferrielectric} $Abm2$ phase. In contrast to
non-degenerate unstable phonons at the $X$ and $R$ points, the unstable phonon
at the $Z$ point is doubly degenerate, and its condensation can lead to two
antiferroelectric phases, $Pmma$ (Fig.~\ref{fig5},\emph{d}) and $Cmcm$, with
displacements in one or both ...--Ti--O--... chains, respectively. These
displacements result in the in-plane microscopic polarization directed along
[100] and [110]. It was the latter of these two phases which was considered
above in the calculation of the interlayer interaction energy.

\begin{table}
\caption{\label{table3}Energies $E$ (in meV) of the antiferroelectric phases
arising from the condensation of unstable phonons at the boundary of the
Brillouin zone and of the ferroelectric $Amm2$ phase for free-standing
(BaTiO$_3$)$_1$(BaZrO$_3$)$_n$ superlattices with $n = 1$, 2, and 3. The energy
of the paraelectric $P4/mmm$ is taken as the energy reference.}
\begin{ruledtabular}
\begin{tabular}{ccccccc}
Unstable & \multicolumn{2}{c}{$n = 1$} & \multicolumn{2}{c}{$n = 2$} & \multicolumn{2}{c}{$n = 3$} \\
\cline{2-7}
phonon   & Phase    & $E$       & Phase  & $E$       & Phase  & $E$      \\
\hline
$X$      & $Pmma$   & $-$37.3   & $Pmma$ & $-$52.1   & $Pmma$ & $-$59.8  \\
         & $Pmc2_1$ & $-$68.1   & $Pm$   & $-$95.2   & $Pm$   & $-$109.0 \\
$R$      & $Cmma$   & $-$33.7   & $Cmma$ & $-$51.1   & $Cmma$ & $-$59.7  \\
         & $Abm2$   & $-$64.9   & $Abm2$ & $-$94.3   & $Abm2$ & $-$108.8 \\
$Z$      & $Pmma$   & $-$41.9   & $Pmma$ & $-$60.7   & $Pmma$ & $-$69.6  \\
         & $Cmcm$   & $-$68.9   & $Cmcm$ & $-$100.3  & $Cmcm$ & $-$114.9 \\
$\Gamma$ & $Amm2$   & $-$72.7   & $Amm2$ & $-$101.0  & $Amm2$ & $-$115.1 \\
\end{tabular}
\end{ruledtabular}
\end{table}

The ferroelectric instability of the intermediate $Pmma$ and $Cmma$ phases
is easily understandable: in these two phases, the structural relaxation
occurs in only one of two ferroelectrically unstable ...--Ti--O--... chains
propagating in the [100] and [010] directions, and so the structure retains
the ferroelectric instability in the second chain. In the $Pmc2_1$, $Pm$,
$Abm2$, and $Cmcm$ phases, the structural relaxation occurs simultaneously
in both chains; this is why these structures are stable (metastable). In all
cases, the obtained intermediate and \emph{ferrielectric} phases have higher
energies in comparison with the ferroelectric $Amm2$ phase (Table~\ref{table3}).

Now we discuss the structure of rotations in the SLs under consideration.
A comparison of the frequencies of unstable phonons corresponding to various
combinations of rotations in the layers of SLs with the same period shows
that phonons with out-of-phase rotations in neighboring layers are
energetically least stable. These rotations are fully consistent with the
instability at the $R$ point in the parent cubic BaZrO$_3$ (Fig.~\ref{fig1}).
The main difference between the two rotation systems corresponding to unstable
phonons at the $M$ and $A$ points is the relation between rotations in the
\emph{neighboring periods} of the superlattice: for the phonon condensation
at the $M$ point, when the structure is translated by one SL period, the
octahedra are rotated in phase; in the case of the phonon condensation at
the $A$ point, they are rotated out of phase. It can be seen that the ground
state of the SL always corresponds to such a rotation system in which the
octahedra are rotated out of phase in any pair of neighboring layers, i.e.,
the $M$ phonon is condensed in the SLs with even period and the $A$ phonon
is condensed in the SLs with odd period.

One more subject for discussion is the comparison of the obtained results with
the experimental data of Ref.~\cite{JApplPhys.108.084104}. According to the
X-ray data from this work, the spontaneous polarization in SLs is predominantly
oriented in the layer plane; however, the line narrowing and an increase in
the frequency of the soft $E$ mode in Raman spectra indicate the monoclinic
distortion of the structure and the appearance of the $z$-component of
polarization in BaTiO$_3$ layers. The $z$-component of polarization determined
from the hysteresis loops nonmonotonically depends on the superlattice period.
The authors supposed that the appearance of the polarization component normal
to the film is caused by the ferroelectric phase transition predicted in
Ref.~\onlinecite{PhysRevB.72.144101} in biaxially compressed BaZrO$_3$ layers.

The calculations performed in the present work showed that the $A_{2u}$
phonon is indeed unstable in biaxially compressed BaZrO$_3$ (space group
$P4/mmm$) with the in-plane lattice parameter equal to the lattice parameter
of the free-standing (BaTiO$_3$)$_1$(BaZrO$_3$)$_1$ SL ($a_0 = 7.7182$~Bohr)
(according to our data, the critical lattice parameter at which the instability
of this phonon appears is $\approx$7.735 Bohr and notably exceeds the value of
7.425 Bohr reported in Ref.~\onlinecite{PhysRevB.72.144101}). However, the
instability of the $A_{2u}$ phonon in biaxially stretched BaTiO$_3$ with the
lattice parameter equal to $a_0$ is absent (the phonon frequency is 45~cm$^{-1}$).
Therefore, the appearing of the $z$-component of polarization in SLs needs
a more detailed analysis.

As we have shown previously,~\cite{PhysSolidState.52.1448}  the $A_{2u}$ mode
(which is responsible for the appearance of the $z$-component of polarization)
is stable (37~cm$^{-1}$) in the paraelectric $P4/mmm$ phase of the free-standing
(BaTiO$_3$)$_1$(BaZrO$_3$)$_1$ SL. The calculations of the phonon frequencies
for the $P4/mmm$ phase of (BaTiO$_3$)$_n$(BaZrO$_3$)$_n$ SLs with $n = 2$, 3,
and 4 performed in this work showed that the frequency of this mode rapidly
decreases with increasing $n$ (28.5~cm$^{-1}$ for $n = 2$ and 17~cm$^{-1}$ for
$n = 3$) and becomes unstable at $n = 4$ (7$i$~cm$^{-1}$). However, the
instability of the $A_{2u}$ phonon in the paraelectric phase does not mean
that the $z$-component of polarization will appear in the ferroelectric ground
state. The in-plane spontaneous polarization, which appears due to the strong
instability of the $E_u$ phonon, abruptly increases the frequency of the
corresponding mode ($B_1$ or $B_2$) in the $Amm2$ phase, and this phase becomes
stable with respect to the appearance of the $z$-component of polarization.

This stability reduction of the $A_{2u}$ phonon with increasing $n$ recalls
the results obtained for the (BaTiO$_3$)$_m$(BaO)$_n$ Ruddlesden--Popper
SLs.~\cite{PhysRevB.82.045426}  In these SLs, a similar decrease in the $A_{2u}$
phonon frequency with increasing the BaTiO$_3$ layer thickness was observed;
at $m \ge 8$, it became unstable. However, already at $m = 4$, the instability
at the $X$ point appearing in the SLs resulted in the formation of
domain-like structure, since it was energetically more favorable to form
a structure with a $z$-component of polarization periodically varying in space,
rather than to uniformly polarize the BaO layer having a low dielectric constant.
The macroscopic $z$-component of polarization is absent in such a structure.

Another possible cause for the appearance of the $z$-component of polarization
can be a substrate-induced strain in the superlattice (in
Ref.~\onlinecite{JApplPhys.108.084104}, the SLs were grown on MgO substrate).
Calculations show that in (BaTiO$_3$)$_n$(BaZrO$_3$)$_n$ SLs with $n = 1$ and
2, for the in-plane lattice parameter equal to $a = 0.99a_0$ and 0.98$a_0$
($a_0$ is the in-plane lattice parameter of the free-standing SL) and the
disabled octahedra rotations, the polarization vector is rotated by $\sim$38$^\circ$
and $\sim$67$^\circ$ from the layer plane (the space group of the ferroelectric
ground state is $Cm$); for $a = 0.97a_0$ and disabled rotations, the vector
is rotated by 90$^\circ$ (the space group of the ferroelectric ground state is
$P4mm$). However, when the octahedra rotations are enabled, in all SLs we see
an unexpected result: the $z$-component of polarization disappears in SLs with
$a = 0.99a_0$ and 0.98$a_0$ (space group of the ground state is $Pmc2_1$ for
$n = 1$ and $Pnc2$ for $n = 2$), whereas in the SL with $a = 0.97a_0$, the
polarization is rotated toward the layer plane by $\sim$24$^\circ$ (the space
group of the ground state is $Pc$).

The competition between the ferroelectric and structural instabilities in
crystals with the perovskite structure has been known for a long
time;~\cite{PhysRevLett.74.2587,Ferroelectrics.206.181}  however, the fact
that the octahedra rotations can have such a strong effect on the polarization
seems to be observed for the first time. This finding can be important for
the following reason. In the SLs under consideration, the energy gain
resulting from the structural distortions is less than the energy gain
resulting from the ferroelectric ordering (compare the energies $\Delta E_1$
and $\Delta E_2$ in Table~\ref{table2}). This leads us to expect that, as
the temperature increases, the structural phase transition in the SLs will
occur at a lower temperature than the ferroelectric phase transition. Then,
since the disappearance of the octahedra rotations results in the onset of
the $z$-component of polarization, the situation is possible that an improper
ferroelectric phase transition accompanied by the appearance of a nonzero
$z$-component of polarization can occur in the SLs when the structural phase
transition temperature is approached from below (when the octahedra rotation
angle rapidly decreases). The abrupt rotation of the polarization vector as
the temperature varies can result in the appearance of a number of physical
property anomalies resembling the anomaly in the piezoelectric coefficients
near the morphotropic phase boundary. However, currently, it is not clear
whether the structural phase transition occurs in BaZrO$_3$. Nevertheless,
even if the establishment of the long-range order in the octahedra rotations
in BaZrO$_3$ is impossible for some reason, the changes in the local angles
of the rotations with temperature can have a significant effect on the
$z$-component of polarization.

The above-considered attempts to explain the appearance of the $z$-component
of polarization, which were based on the assumption of a uniformly strained
BaTiO$_3$/BaZrO$_3$ SL, were not successful (the strong biaxial substrate-induced
in-plane compression of the SL cannot be retained in a thick SL). It seems that
the only way to explain this phenomenon is to consider the stress relaxations
occurring in SLs with thick layers. This idea has already been used to explain
the appearance of the $z$-component of polarization in BaTiO$_3$/BaZrO$_3$ SLs
with large periods.~\cite{ApplPhysLett.92.102903,PhysSolidState.52.1448}
Our calculations show that the critical in-plane lattice parameter at which
the $z$-component of polarization appears in BaTiO$_3$ is $\sim$7.565~Bohr;
in this case, the out-of-plane lattice parameter is $c \approx 7.471$~Bohr.
If we take into account the systematic error ($-$0.7\%) in the prediction of
the lattice parameters for BaTiO$_3$, the experimental parameter $c = 3.981$~{\AA}
in the BaTiO$_3$ layer should correspond to this situation. According to the
X-ray data,~\cite{JApplPhys.108.084104}  such a relaxation of mechanical stress
already appears in the SL with a period of $L = 32$~{\AA}. So, this relaxation
explains the appearance of the $z$-component of polarization in SLs with the
specified and larger periods.

The first-principles calculations allow us to understand why the quality of
short-period BaTiO$_3$/BaZrO$_3$ and BaTiO$_3$/Ba(Ti,Zr)O$_3$ SLs grown in
the [001] direction is not very good.~\cite{PhysRevB.74.064107,JApplPhys.108.084104}
This is because of the tendency of BaTi$_{1-x}$Zr$_x$TiO$_3$ solid solutions
to three-dimensional chessboard-type ordering of cations at the $B$ sites.
The calculations show that the energy of the paraelectric phase of the
(BaTiO$_3$)$_1$(BaZrO$_3$)$_1$ SL with the elpasolite structure (space group
$Fm3m$), which appears when growing the SL in the [111] direction, is by 83~meV
(per formula unit) lower than the energy of the same SL but grown in the
[001] direction. It is possible that defects generated during the stress
relaxation in SLs grown in the [001] direction become the nucleation centers
of three-dimensional chessboard-type ordered inclusions. Furthermore, the
possible electrical activity of these defects can be a factor which changes
the electrostatic boundary conditions at the heterointerface and stabilizes
the polar state in the BaTiO$_3$ layer without inducing an appreciable electric
field and polarization in the BaZrO$_3$ layer.

\section{Conclusion}

The properties of BaTiO$_3$/BaZrO$_3$ superlattices with competing ferroelectric
and structural instabilities were calculated from first principles within the
density functional theory. It was established that the quasi-two-dimensional
ferroelectricity with the polarization oriented in the layer plane, which weakly
interacts with the polarization in neighboring layers, occurs in SLs grown in
the [001] direction with one-unit-cell thick barium titanate layer. It was shown
that the energy gain from the ferroelectric ordering and the height of the
potential barrier separating different orientational states of polarization
are sufficiently large to observe the formation of an array of independent
polarized planes at room temperature. The effect of the structural instability
on the properties of SLs was considered. It was shown that the ground state is
a result of simultaneous condensation of the ferroelectric $\Gamma_{15}$ mode
and phonons at the $M$ point (for SLs with even period) or at the $A$ point (for
SLs with odd period). Thus, the ground state is a structure with out-of-phase
rotations in neighboring layers, in which highly polarized layers and layers
with strong octahedra rotations are spatially separated. This significantly
reduces the influence of structural distortions on the spontaneous polarization.
It was shown that the switching on of the octahedra rotations in biaxially
compressed BaTiO$_3$/BaZrO$_3$ SLs with an in-plane strain of 1--3\% results in
an abrupt change in the polarization direction. This suggests that an improper
ferroelectric phase transition accompanied by the appearance of a nonzero
$z$-component of polarization can occur in these SLs with increasing temperature.

\begin{acknowledgments}
The calculations presented in this work were performed on the laboratory
computer cluster (16 cores) and the SKIF-MGU (``Chebyshev'') and ``Lomonosov''
supercomputers.
\end{acknowledgments}

\providecommand{\BIBYu}{Yu}

\end{document}